\documentclass[a4paper,11pt]{article}
\usepackage{jheppub} % for details on the use of the package, please
\usepackage[T1]{fontenc} % if needed
\usepackage{subfigure}
\graphicspath{{Figures/}}

\title{\boldmath Scalar-gravitational quasinormal modes and echoes in a five dimensional thick brane}

\author[a,b]{Weike Deng,}
\author[c]{Sheng Long,}
\author[b,1]{Qin Tan }
\author[b,1]{Zu-Cheng Chen,}
\author[b,1]{and Jiliang Jing\note{Corresponding author.}}

\affiliation[a]{School of Science, Hunan Institute of Technology, Hengyang 421002, P. R. China}
\affiliation[b]{Department of Physics, Key Laboratory of Low Dimensional Quantum Structures and Quantum Control of Ministry of Education, Synergetic Innovation Center for Quantum Effects and Applications, Hunan Normal University, Changsha, 410081, Hunan, China}
\affiliation[c]{School of Fundamental Physics and Mathematical Sciences, Hangzhou Institute for Advanced Study, University of Chinese Academy of Sciences, Hangzhou 310024, China}

\emailAdd{wkdeng@hnit.edu.cn}
\emailAdd{shenglong@ucas.ac.cn}
\emailAdd{zuchengchen@hunnu.edu.cn}
\emailAdd{tanqin@hunnu.edu.cn}
\emailAdd{jljing@hunnu.edu.cn}

\abstract{The scalar perturbations of thick braneworld models provide critical insights into their matter-geometry relationship, distinct from tensor modes. This work systematically investigates quasinormal modes and gravitational echoes from scalar perturbations in a thick brane model exhibiting internal structure and brane splitting. Using the WKB method, direct integration, and Bernstein spectral techniques, we compute quasinormal frequencies across different parameter regimes, addressing both single and double‐barrier effective potentials. Time-domain evolution of wave packets reveals clear echo signals for split brane configurations ($s > 1, \delta > 1$), produced by successive reflections between sub-branes. A key finding is the position-dependence of echo modes within the extra dimension: observers located on a sub-brane detect clean periodic signals, whereas those situated between sub-branes observe more complex, modulated waveforms. This effect offers a distinct signature of the brane’s internal structure. The observed echoes, along with consistent frequency- and time-domain results, advance the understanding of thick brane dynamics and open an observational window into warped extra dimensions. Moreover, the similarity between the effective potential in thick brane scenarios and those of black holes and wormholes offers valuable perspectives for studying echo-related phenomena in these gravitational systems.}

\begin{document} 
\maketitle
\flushbottom

\section{Introduction}
Since the 20th century, the concept that our observable universe could be a four-dimensional hypersurface embedded within a higher-dimensional spacetime has profoundly influenced theoretical physics, offering novel perspectives on the fundamental structure of reality. This idea, initially motivated by the quest to unify fundamental interactions—exemplified by Kaluza-Klein (KK) theory and string theory~\cite{kaluza:1921un,Klein:1926tv,Scherk:1974ca}—has evolved into a broad framework for addressing some of the most persistent problems in modern physics. Among these developments, the Randall-Sundrum (RS) models stand out for their innovative approach to resolving the hierarchy problem between the electroweak and Planck scales~\cite{Randall:1999ee,Randall:1999vf}. The RS-II model, in particular, demonstrated that an infinite extra dimension is compatible with four-dimensional gravity localized on a brane, sparking extensive research into its implications for cosmology, black hole physics, and phenomenology beyond the Standard Model~\cite{Shiromizu:1999wj,Tanaka:2002rb,Gregory:2008rf,Konoplya:2011qq,Bauer:2016lbe,Jaman:2018ucm,Adhikari:2020xcg,Bhattacharya:2021jrn,Geng:2020fxl,Geng:2021iyq,Geng:2022dua}.

A natural and physically rich generalization of the RS-II scenario is the thick brane construct, wherein the brane is generated dynamically by one or more bulk matter fields, such as a scalar field, leading to a smooth energy density distribution along the extra dimension~\cite{DeWolfe:1999cp,Gremm:1999pj,Csaki:2000fc}. Unlike their infinitely thin counterparts, thick branes possess finite thickness and internal structure, which significantly enriches their phenomenology. Considerable effort has been devoted to constructing thick brane solutions across various gravitational theories and examining the localization properties of both the gravitational zero mode and diverse matter fields~\cite{Afonso:2007gc,Dzhunushaliev:2010fqo,Dzhunushaliev:2011mm,Geng:2015kvs,Melfo2006,Almeida2009,Zhao2010,Chumbes2011,Liu2011,Bazeia:2013uva,Xie2017,Gu2017,ZhongYuan2017,ZhongYuan2017b,Zhou2018,Hendi:2020qkk,Xie:2021ayr,Moreira:2021uod,Xu:2022ori,Silva:2022pfd,Xu:2022gth,Liu:2017gcn,Ahluwalia:2022ttu}. While the localization of zero modes is essential for recovering standard four-dimensional physics on the brane, it is the massive KK modes—carrying information about the extra-dimensional geometry—that offer a window into new physics. Recent studies have revealed that, within the seemingly continuous spectrum of massive KK modes, there exists a discrete set of quasinormal modes (QNMs)~\cite{Seahra:2005wk,Seahra:2005iq,Tan:2022vfe,Tan:2023cra,Jia:2024pdk,Tan:2024url,Tan:2024aym,Jia:2024sdk}, whose spectrum is intrinsically tied to the brane's structure.

The complete gravitational fluctuations of a thick brane are composed of tensor, vector, and scalar modes~\cite{Giovannini:2001fh,Giovannini:2001xg,Giovannini:2001vt,Kobayashi:2001jd}. Previous investigations into brane QNMs have predominantly focused on transverse-traceless tensor fluctuations, corresponding to spin-2 gravitons. However, for thick branes formed by coupled scalar-gravity systems, scalar fluctuations (graviscalars) are an indispensable part of the spectrum, encoding details of the matter configuration generating the brane. Scalar perturbations have indeed been studied in diverse higher-dimensional contexts, including brane cosmology~\cite{Langlois:2000iu,Koyama:2004ap,Hiramatsu:2004aa,Brax:2004xh,Maartens:2010ar,Maier:2013gua,Jaman:2018ucm,Banerjee:2020uil,Ravanpak:2022awx} and higher-dimensional black holes~\cite{Cardoso:2003vt,Natario:2004jd,Konoplya:2008ix}. A particularly intriguing feature within certain thick brane models is the phenomenon of brane splitting, where a single brane separates into multiple sub-branes~\cite{Melfo:2002wd,Castillo-Felisola:2004omi,deBrito:2014pqa,deSouzaDutra:2014ddw,Farokhtabar:2016fhm,Xie:2019jkq}. This splitting not only provides a novel mechanism for addressing the hierarchy problem within the RS-II framework~\cite{Guerrero:2006gj,Ahmed:2012nh,deSouzaDutra:2013rwa} but also creates a complex, multi-barrier effective potential for perturbations. Such a potential is a known prerequisite in various gravitational contexts for the emergence of delayed, repeating signals known as echoes~\cite{Cardoso:2017cqb,Cardoso:2019rvt,Mark:2017dnq,Conklin:2017lwb,Barcelo:2017lnx,Qian:2024zvq,Lin:2023qgd}, suggesting that split thick branes could be a natural arena for this phenomenon~\cite{Zhu:2024gvl,Tan:2024qij}.

This paper presents a systematic study of the quasinormal modes and the potential echo signals associated with scalar fluctuations in a thick brane model exhibiting splitting behavior. We employ a multi-faceted approach: in the frequency domain, we calculate the QNM spectra using the WKB approximation, the direct integration method, and the Bernstein spectral method, adapting to the challenges posed by different parameter regimes. In the time domain, we numerically evolve initial wave packets to directly observe the ringdown waveforms and identify clear echo signals for specific parameter choices. Our analysis reveals that the echo signature is highly sensitive to the observer's position within the extra dimension, providing a unique and potentially detectable fingerprint of the brane's internal structure. These findings not only advance our understanding of braneworld dynamics but also offer valuable insights and methodological approaches for investigating similar echo phenomena in black hole and wormhole spacetimes.

The paper is organized as follows. In Sec.~\ref{sec:background}, we review the thick brane model generated by a canonical scalar field and derive the master equation for scalar perturbations. Section~\ref{sec:qnm_echo} is devoted to our main analysis: we compute the QNM frequencies in the frequency domain and then investigate the full dynamical response, including echoes, through time evolution simulations. Finally, we conclude and discuss the implications of our findings in Sec.~\ref{sec:conclusions}.

\section{Braneworld model in general relativity}
\label{sec:background}
In this section, we briefly review the thick brane solution in five-dimensional general relativity and obtain the fluctuation equation for the scalar gravitational perturbation. A thick brane may arise from different bulk fields, such as scalars or vectors. We choose that the thick brane is generated by a canonical scalar field. Here we adopt a canonical scalar as the source. The action is
\begin{eqnarray}
	S=\int d^5x\sqrt{-g}\left(\frac{1}{2\kappa^{2}_{5}}R-\frac{1}{2}g^{MN}\partial_{M}
	\varphi\partial_{N}\varphi -V(\varphi)\right),\label{action}
\end{eqnarray}
with $\kappa_{5}^{2}=8\pi G_{5}$. Throughout, we set $\kappa_{5}= 1$. Varying the action yields the Einstein equations and the scalar field equation,
\begin{eqnarray}
	R_{MN}-\frac{1}{2}Rg_{MN}&=&-\frac{1}{2}g_{MN}\left(\partial^{A}\varphi\partial_{A}\varphi-V(\varphi)\right) \nonumber\\
	&&+\partial_{M}\varphi\partial_{N}\varphi,\label{field equation}\\
	g^{MN}\nabla_{M}\nabla_{N}\varphi&=&\frac{\partial V(\varphi)}{\partial\varphi}.\label{motion equation}
\end{eqnarray}
Indices $M,N,\dots=0,1,2,3,5$  refer to bulk coordinates; $\mu,\nu\dots=0,1,2,3$ are tangent to the brane; and $i,j\dots=1,2,3$ label spatial directions on the brane.

Working in the so-called gauge coordinate system, a static, Poincaré-invariant brane geometry can be written as~\cite{Melfo:2002wd}
\begin{equation}
	ds^2=e^{2A(y)}\eta_{\mu\nu}dx^\mu dx^\nu+e^{2H(y)}dy^2,
	\label{metric}
\end{equation}
where $\eta_{\mu\nu}=\text{diag}(-1,1,1,1)$. Substituting this ansatz into the above field equations gives the specific dynamical equations
\begin{eqnarray}
	2A'^2-A'H' +A''&=&-\frac{1}{2}\varphi'^2-V,  \label{EoMs1}\\
	6A'^2&=&\frac{1}{2}\varphi'^2-V,  \label{EoMs2}\\
	\varphi{''}+4A'\varphi'&=&\frac{\partial V}{\partial\varphi},  \label{EoMsphi}
\end{eqnarray}
where a prime denotes derivative with respect to $y$. 

A known family of thick-brane solutions is~\cite{Melfo:2002wd}:
\begin{eqnarray}
	A(y)&=&\delta H(y)=\frac{-\delta}{2s}\ln\left[1+\left(\frac{ky}{\delta}\right)^{2s}\right] \label{warpfactorsolution1},\\
	\varphi(y)&=&\frac{\sqrt{3\delta(2s-1)}}{s}\arctan\left(\frac{ky}{\delta}\right),\label{scalarfieldsolution1}\\
	V(\varphi)&=&3k^{2}\left[\frac{2s+4\delta-1}{2\delta}\cos\left(\frac{s\varphi}{\sqrt{3\delta(2s-1)}}\right)^{2}-2\right]\left[\sin\left(\frac{s\varphi}{\sqrt{3\delta(2s-1)}}\right)\right].\label{scalarpotentialsolution1}
\end{eqnarray}
Here, $\delta$ and $s$ are dimensionless parameters controlling the brane thickness, and $k$ is a parameter of mass dimension one. The parameter $s$ is restricted to positive odd integers. These solutions represent generalizations of the thick brane model proposed by Gremm et al.~\cite{Gremm:1999pj}. Specifically, the case $s = \delta = 1$ recovers the solution in Ref.~\cite{Gremm:1999pj}, while the limit $\delta \to 0$ reproduces the RS-II thin brane model.

\begin{figure}
	\centering
	\subfigure[~$\delta=1$]{\label{figw123P1}
		\includegraphics[width=0.4\textwidth]{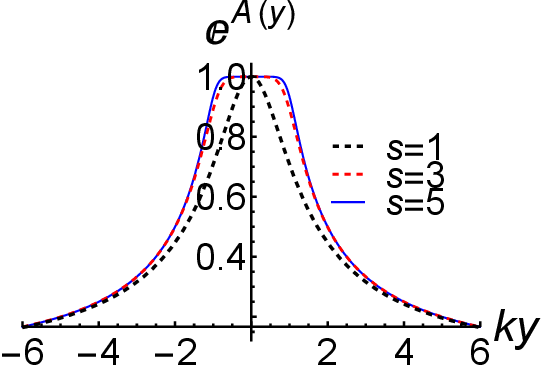}}
	\subfigure[~$s=3$]{\label{figw456P2}
		\includegraphics[width=0.4\textwidth]{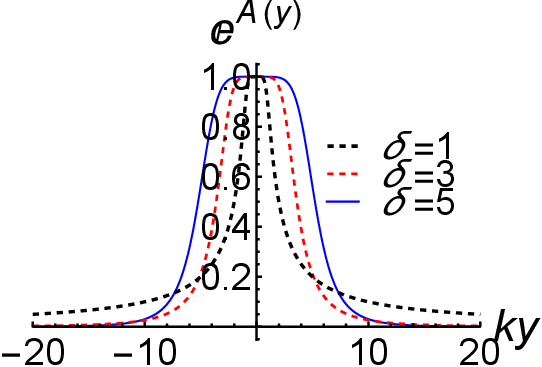}}
	\caption{The shapes of the warp factor~\eqref{warpfactorsolution1}.}\label{figwarpfactor}
\end{figure}

\begin{figure}
	\centering
	\subfigure[~$\delta=1$]{\label{figphiP1}
		\includegraphics[width=0.4\textwidth]{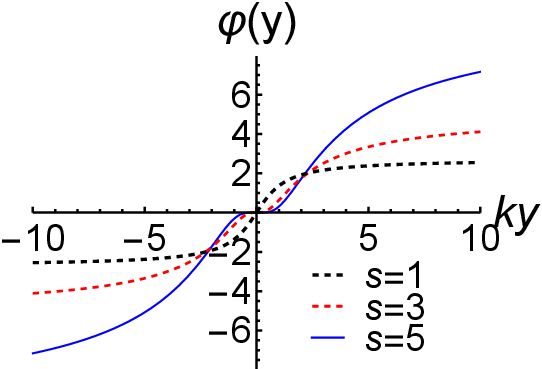}}
	\subfigure[~$s=3$]{\label{figphiP2}
		\includegraphics[width=0.4\textwidth]{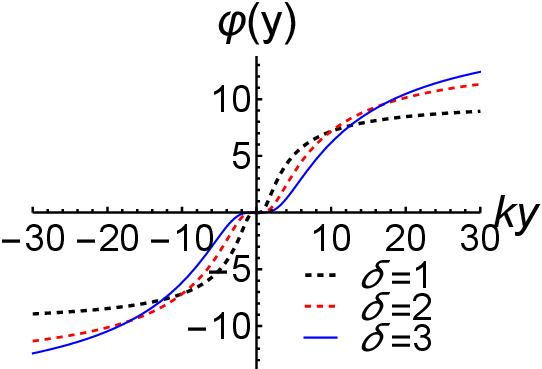}}
	\caption{Plots of the scalar field~\eqref{scalarfieldsolution1}.}\label{figphi}
\end{figure}

\begin{figure}[htbp]
	\centering
	\subfigure[~$\delta=1$]{\label{figVphiP1}
		\includegraphics[width=0.4\textwidth]{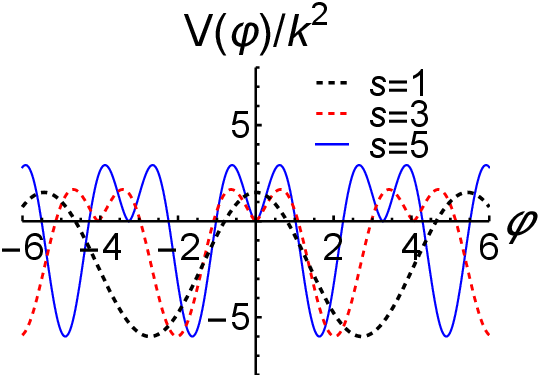}}
	\subfigure[~$s=3$]{\label{figVphiP2}
		\includegraphics[width=0.4\textwidth]{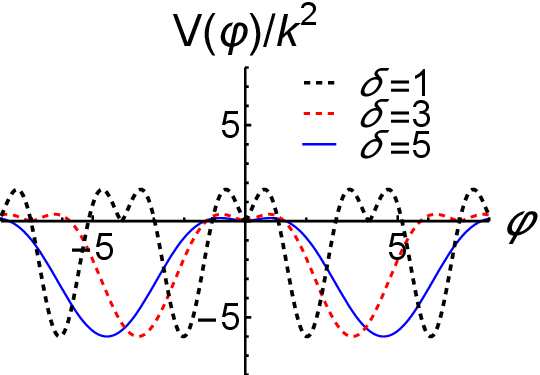}}
	\caption{Plots of the scalar potential~\eqref{scalarpotentialsolution1}.}\label{figVphi}
\end{figure}

\begin{figure}
	\centering
	\subfigure[~$\delta=1$]{\label{figrhoP1}
		\includegraphics[width=0.4\textwidth]{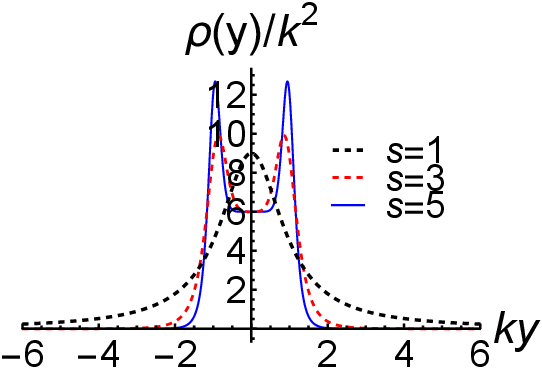}}
	\subfigure[~$s=3$]{\label{figrhoP2}
		\includegraphics[width=0.4\textwidth]{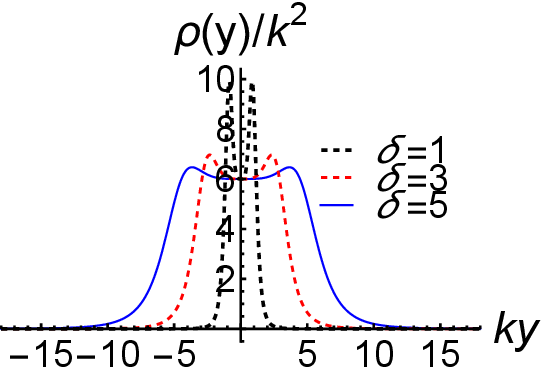}}
	\caption{Plots of the energy density~\eqref{energydenisty}.}\label{figrho1}
\end{figure}

The profiles of the warp factor~\eqref{warpfactorsolution1}, scalar field~\eqref{scalarfieldsolution1}, and scalar potential~\eqref{scalarpotentialsolution1} for various parameter values are illustrated in Figs.~\ref{figwarpfactor},~\ref{figphi}, and~\ref{figVphi}. The warp factor broadens as $s$ or $\delta$ increases. The scalar field profile changes from a single kink for $s = 1$ to a double-kink for $s > 1$. Moreover, as $s$ and $\delta$ increase, the scalar field becomes increasingly flat around $y = 0$. For $s > 1$, the scalar potential exhibits a splitting at $\varphi = 0$, as seen in Fig.~\ref{figVphi}. These characteristics suggest that the thick brane splits into two sub-branes when $s > 1$, with the splitting becoming more pronounced as $s$ and $\delta$ grow. This conclusion is further supported by the energy density distribution:
\begin{eqnarray}
\rho(y)&=&-3 e^{-2 H(y)} \left[A''(y)-A'(y) H'(y)+2 A'(y)^2\right]\nonumber\\
		&=&\frac{-3 \delta }{y^2} \left[\left(\frac{ky}{\delta }\right)^{2 s}+1\right]^{\frac{1}{s}-2} \left(\frac{ky}{\delta }\right)^{2 s}\left[2 \delta  \left(\frac{ky}{\delta }\right)^{2 s}-2 s+1\right]. \label{energydenisty}
\end{eqnarray}
Figure~\ref{figrho1} displays the energy density profiles. Clearly, for $s > 1$, the energy density splits into two distinct peaks, confirming the presence of two sub-branes. The distance between these peaks increases with larger values of $s$ and $\delta$.

To analyze the QNMs of scalar fluctuations on the thick brane, we proceed by introducing scalar perturbations to the metric. In braneworld scenarios, studying metric and field fluctuations typically requires a transition to conformally flat coordinates. We therefore perform the coordinate transformation:
\begin{equation}
	dz=e^{H-A}dy,
\end{equation}
which brings the metric~\eqref{metric} into the form:
\begin{equation}
	ds^2=e^{2A(z)}(\eta_{\mu\nu}dx^\mu dx^\nu+dz^2).
	\label{metric2}
\end{equation}
In linear perturbation theory, the metric fluctuations can be decomposed into scalar, transverse vector, and transverse-traceless tensor modes. These components decouple under a suitable scalar-vector-tensor decomposition, allowing us to focus exclusively on scalar perturbations. In the longitudinal gauge, the perturbed metric is given by~\cite{Kobayashi:2001jd}:
\begin{equation}
	ds^2=e^{2A(z)}(1+2\varphi(x^{\mu},z))(\eta_{\mu\nu}dx^\mu dx^\nu+(1+2\Psi(x^{\mu},z))dz^2),
	\label{perturbedmetric}
\end{equation}
while the background scalar field is perturbed as $\phi_0 = \phi(z) + \delta\phi(x^\mu, z)$. Substituting these expressions into Eqs.~\eqref{field equation} and~\eqref{motion equation} yields the system of scalar perturbation equations:
\begin{eqnarray}
	(z,z):&&3\eta^{\alpha\beta}\partial_{\alpha}\partial_{\beta}\Psi+12\partial_{z}A\partial_{z}\Psi-12(\partial_{z}A)^2\partial_{z}\varphi \nonumber\\ 
	&&=\partial_{z}\phi\partial_{z}\delta\phi-\varphi(\partial_{z}\delta\phi)^2-e^{2A}\frac{\partial V}{\partial \phi}\delta\phi,\label{zzcompont}\\
	(z,\mu):&&-3\partial_{\mu}\partial_{z}\Psi+3\partial_{z}A\partial_{\mu}\varphi=\partial_{z}\phi\partial_{\mu}\delta\phi,\label{zmucompont}\\
	(\mu,\nu):&&\Big(3\partial_{z}^{2}\Psi-6\partial_{z}^{2}A\varphi-3\partial_{z}A\partial_{z}\varphi+9\partial_{z}A\partial_{z}\Psi\nonumber\\&&-6(\partial_{z}A)^{2}\varphi+\eta^{\alpha\beta}\partial_{\alpha}\partial_{\beta}\varphi+2\eta^{\alpha\beta}\partial_{\alpha}\partial_{\beta}\Psi\Big)\delta_{\nu}^{\mu}\nonumber\\
	&&-\eta^{\mu\beta}\partial_{\beta}\partial_{\nu}\varphi-2\eta^{\mu\beta}\partial_{\beta}\partial_{\nu}\Psi\nonumber\\
	&&=\Big(-\partial_{z}\phi\partial_{z}\delta\phi+\varphi(\partial_{z}\delta\phi)^2-e^{2A}\frac{\partial V}{\partial \phi}\delta\phi\Big)\delta_{\nu}^{\mu},\nonumber\\ \label{munucompont}\\
	\text{matter}:&&\partial_{z}^{2}\delta\phi+3\partial_{z}A\partial_{z}\delta\phi+(4\partial_{z}\Psi-\partial_{z}\varphi-6\partial_{z}\varphi)\partial_{z}\phi\nonumber \\ &&-2\varphi\partial_{z}^{2}\phi_{0}+\eta^{\alpha\beta}\partial_{\alpha}\partial_{\beta}\delta\phi=e^{2A}\frac{\partial V}{\partial \phi}\delta\phi.\label{mattercompont}
\end{eqnarray}
From the off-diagonal components of Eq.~\eqref{munucompont}, we deduce:
\begin{equation}
	\varphi+2\Psi=0.\label{offdiagonalmunucompont}
\end{equation}
Equation~\eqref{zmucompont} implies:
\begin{equation}
	\delta\phi=\frac{(-3\partial_{z}\Psi+3\partial_{z}A\varphi)}{\partial_{z}\phi}.\label{zmucompont1}
\end{equation}
indicating that only one physical scalar degree of freedom remains. Combining Eqs.~\eqref{zzcompont}–\eqref{offdiagonalmunucompont}, the master equation for the scalar perturbation is derived as:
\begin{eqnarray}
	\Box^{(4)}\Psi&&+\left(4\partial_{z}^{2}A-\frac{4\partial_{z}A\partial_{z}^{2}\phi}{\partial_{z}\phi}\right)\Psi+\left(3\partial_{z}A-
	\frac{2\partial_{z}^{2}\phi}{\partial_{z}\phi}\right)\partial_{z}\Psi+\partial_{z}^{2}\Psi=0, \label{mainescalarquation}
\end{eqnarray}
where $\Box^{(4)} = \eta^{\alpha\beta}\partial_{\alpha}\partial_{\beta}$. We now perform a Kaluza-Klein decomposition:
\begin{equation}
	\Psi(x^{\mu},z)=e^{-\frac{3}{2}A(z)}\partial_{z}\phi \tilde{\Psi}(t,z) e^{-i a_{i}x^{i}},\label{decomposition1}
\end{equation}
leading to the wave equation:
\begin{equation}
	-\partial_{t}^{2}\tilde{\Psi}+\partial_{z}^{2}\tilde{\Psi}-U(z)\tilde{\Psi}-a^{2}\tilde{\Psi}=0, \label{evolutionequation}
\end{equation}
with the effective potential given by:
\begin{eqnarray}
	U(z)&=&-\frac{5}{2}\partial_{z}^{2}A+\frac{9}{4}(\partial_{z}A)^{2}-\frac{\partial_{z}^{3}\phi}{\partial_{z}\phi}+\partial_{z}A\frac{\partial_{z}^{2}\phi}{\partial_{z}\phi}+2\left(\frac{\partial_{z}^{2}\phi}{\partial_{z}\phi}\right)^{2}\label{effectivepotential}
\end{eqnarray} 
and $a = \sqrt{a^i a_i}$ representing the magnitude of the spatial momentum on the brane. When $a = 0$, the Kaluza-Klein (KK) mode propagates solely along the extra dimension, moving at the speed of light asymptotically. For $a > 0$, the mode possesses both extra-dimensional and brane-directed motion. Further decomposing $\tilde{\Psi} = e^{-i\omega t} \psi(z)$ and substituting into Eq.~\eqref{evolutionequation} yields a Schrödinger-like equation:
\begin{equation}
	-\partial_{z}^{2}\psi(z)+U(z)\psi(z)=m^{2}\psi(z),\label{Schrodingerlikeequation}
\end{equation}
where $m = \sqrt{\omega^2 - a^2}$ is the mass of the graviscalar KK mode. In contrast to tensor perturbations, a bound zero mode in the scalar sector is undesirable, as it would introduce a fifth force on the brane. To examine the zero mode, we express the equation in supersymmetric form:
\begin{equation}
	H^{\dagger}H\phi(z)=m^{2}\psi(z)\label{supersymmetricform},
\end{equation}
with the operators defined as:
\begin{eqnarray}
	H^{\dagger}&=&-\partial_{z}+\frac{3}{2}\partial_{z}A(z)-\frac{\partial^{2}_{z}A(z)}{\partial_{z}A(z)}+\frac{\partial^{2}_{z}\phi(z)}{\partial_{z}\phi(z)},\\
	H&=&\partial_{z}+\frac{3}{2}\partial_{z}A(z)-\frac{\partial^{2}_{z}A(z)}{\partial_{z}A(z)}+\frac{\partial^{2}_{z}\phi(z)}{\partial_{z}\phi(z)}.
\end{eqnarray}
The zero-mode solution is $\psi_0 \propto \dfrac{\partial_z A}{e^{\frac{3}{2}A} \partial_z \phi}$, which is generally non-normalizable and thus not localized on the brane. Beyond the zero mode, the spectrum includes massive KK modes. Although the complexity of $U(z)$ precludes exact analytical solutions for these modes, their behavior can be effectively studied through the framework of quasinormal modes.

\section{quasinormal modes and echo of the branes}
\label{sec:qnm_echo}

In this section, we conduct a comprehensive investigation into the QNMs and the echo signals produced by the thick brane system. To determine the QNMs, we employ multiple complementary methods: the WKB approximation~\cite{Iyer:1986np, Kanti:2006ua,Konoplya:2003ii}, the direct integration method~\cite{Pani:2013pma}, and the Bernstein spectral method~\cite{Fortuna:2020obg}. Furthermore, we explore the emergence of gravitational echoes through numerical time evolution simulations.

\subsection{Frequency domain: QNFs of thick brane}
We begin by examining the graviscalar quasinormal modes within the frequency domain. A notable challenge arises when the parameter $\delta > 1$, as the coordinate transformation between the original $y$ coordinate and the conformal $z$ coordinate becomes analytically intractable. Fortunately, the WKB approximation and the direct integration method allow us to compute the quasinormal frequencies directly in the $y$ coordinate. We therefore categorize our analysis into three distinct parametric regimes:
\begin{enumerate}
	\centering
	\item Case 1: $s = 1$, $\delta \geq 1$
	\item Case 2: $\delta = 1$, $s \geq 1$
	\item Case 3: $s > 1$, $\delta > 1$
\end{enumerate}
Different numerical techniques are applied to compute the quasinormal frequencies in each case, as appropriate.

\subsubsection{Case 1: $s = 1, \delta\geq 1$}
\begin{figure}
	\centering
		\includegraphics[width=0.4\textwidth]{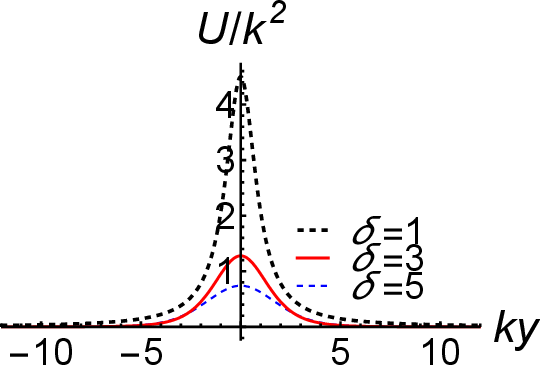}
	\caption{Plots of the effective potential $U(y)$ with $s=1$.}\label{figeffectivePcase1}
\end{figure}

For $s = 1$, the thick brane retains a single, non-splitting profile. However, increasing $\delta$ results in a gradual widening of the brane's thickness. The relationship between $y$ and $z$ is non-trivial for $\delta \neq 1$, prompting us to solve the Schr\"{o}dinger-like equation directly in the $y$ coordinate:
\begin{equation}
e^{2 A(y)-2 H(y)} \left((A'(y)-H'(y)) \psi '(y)+\psi ''(y)\right)+ \left(m ^2-U(y)\right)\psi (y)=0,\label{Schrodingerlikeequation_y}
\end{equation}
where $U(y)$ denotes the effective potential in the $y$-coordinate. The profile of $U(y)$, depicted in Fig.~\ref{figeffectivePcase1}, exhibits a pure potential barrier that vanishes asymptotically. Such a potential does not support bound states but allows for a spectrum of quasinormal modes with complex frequencies, indicative of finite lifetimes. We will now study these QNMs. For Eq.~\eqref{Schrodingerlikeequation_y}, the boundary conditions for the QNMs are unclear. However, noting that Eq.~\eqref{Schrodingerlikeequation_y} is derived from the Schrödinger equation  ~\eqref{Schrodingerlikeequation} transformation, the boundary conditions can be simply written as:
\begin{equation}
	\label{boundaryconditionscase1}
	\psi(z) \propto \left\{
	\begin{aligned}
		e^{im z(y)}, &~~~~~z(y)\to\infty.& \\
		e^{-im z(y)},  &~~~~~z(y)\to-\infty,&
	\end{aligned}
	\right.
\end{equation}
With the boundary conditions and equations in place, we can use various methods to solve the quasinormal frequencies (QNFs). Given the barrier-like nature of the potential, the WKB approximation is well-suited for calculating the fundamental and low-overtone modes.

\textbf{WKB approximation}
The WKB method, widely used in black hole perturbation theory~\cite{Iyer:1986np, Kanti:2006ua}, matches asymptotic WKB solutions at infinity to a Taylor expansion near the potential peak. The sixth-order WKB formula for the complex frequency is given by~\cite{Konoplya:2003ii}:
\begin{eqnarray}
	i\frac{m^{2}-U_{0}}{\sqrt{-2U''_{0}}}-\sum_{j=2}^{6}\Lambda_{j}=n+\frac{1}{2},~~~n=1,~2~,3\dots,\label{WKBm}
\end{eqnarray}
where $U_0$ is the maximum value of $U(z)$, $U''_0$ is the second derivative at the maximum, and $\Lambda_j$ are higher-order correction terms. The explicit form of $\Lambda_{j}$ can be see in Ref.~\cite{Konoplya:2003ii}. Using the WKB method, we obtain some scalar QNFs of the brane, which can be seen from Tab.~\ref{tab1}. For higher overtones, the QNFs can not be solved by the WKB method. To verify the validity of our results, we also recalculated using the direct integration method.

\textbf{Direct integration method} To validate the WKB results, we also employ the direct integration method~\cite{Pani:2013pma}. This method involves integrating the differential equation from both infinities to a matching point, imposing the boundary conditions~\eqref{boundaryconditionscase1}, and requiring continuity of the wave function and its derivative. It should be noted that when the magnitudes of the real and imaginary parts of the QNFs are relatively small, the direct integration method is no longer applicable. Therefore, this method can only solve the frequencies of the fundamental mode and a few low overtones. Table~\ref{tab1} presents the results obtained by using the WKB approximation and the direct integration method. We found that for the fundamental mode and lower overtone modes, the results obtained by the two methods are highly consistent, which proves that our results are valid.

\begin{table}[htbp]
	\centering
		\begin{tabular}{|c|c|c|c|c|}
		\hline
		$\;\delta\;$  &
		$\;n\;$  &
		$\text{WKB method}$  &
		$\;\;\;\text{Direct integration method}\;\;\;$ 	&
		$\;\;\;\text{Time evolution}\;\;\;$\\
		\hline
		~  &~   &$\text{Re}(m/k)$  ~~  $\text{Im}(m/k)$  &$\text{Re}(m/k)$ ~ $\text{Im}(m/k)$  &$\text{Re}(m/k)$ ~ $\text{Im}(m/k)$     \\
		1    &1   &1.99963~~ -0.50721           &~~1.99929 ~~ -0.50734               &~~1.99683~~~~   -0.50018\\
	       	&2   &1.75237~~ -1.57306           &~~1.74655 ~~ -1.57582               &~~ non~~~~~~~~non \\
		2    &1   &1.34122 ~~ -0.33752           &~~1.34121 ~~ -0.33756               &~~1.34089~~~~   -0.33764\\
		&2   &1.21162~~ -1.04621           &~~1.13342 ~~ -1.04160               &~~ non~~~~~~~~non \\
		3    &1   &1.07451~~ -0.26765           &~~1.07451  ~~ -0.26765               &~~1.07459~~~~   -0.26766\\
		4    &1   &0.92147~~ -0.22787           &~~0.91586 ~~ -0.22979               &~~0.92144~~~~   -0.22791\\
		\hline
	\end{tabular}
	\caption{Quasinormal frequencies for Case 1 ($s=1$) obtained via the WKB approximation, direct integration, and time evolution methods. ``non'' indicates modes not extractable from the time domain signal.\label{tab1}}
\end{table}

\subsubsection{Case 2:  $\delta = 1, s\geq1$}
\begin{figure}
	\centering
	\includegraphics[width=0.4\textwidth]{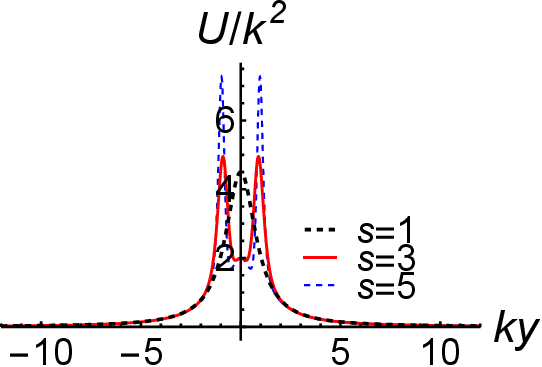}
	\caption{Plots of the effective potential $U(y)$ with $\delta=1$.}\label{figeffectivePcase2}
\end{figure}

For $s=1$, the brane is unsplit and the effective potential remains a single barrier. However, for $s > 1$, the potential splits into a double-barrier structure, as shown in Fig.~\ref{figeffectivePcase2}. This renders the standard WKB approximation less effective. To accurately determine the QNF spectrum, including higher overtones, we utilize the Bernstein spectral method alongside the direct integration method.

\textbf{Bernstein spectral method} This powerful numerical technique solves eigenvalue problems for ordinary differential equations by expanding the solution in a basis of Bernstein polynomials~\cite{Fortuna:2020obg}. Consider a general linear differential operator $\hat{L}(u, \omega)$:
\begin{equation}
	\hat{L}(u, \omega) \Phi(u) = 0, \quad u \in [a, b],
	\label{lineareq}
\end{equation}
where $\omega$ is the eigenvalue. The solution is expanded as:
\begin{equation}
	\Phi(u) = \sum_{k=0}^{N} C_k B_k^N(u),
	\label{basisexpand}
\end{equation}
where $B_k^N(u)$ are Bernstein polynomials. Substituting this expansion into Eq.~\eqref{lineareq} leads to a generalized eigenvalue problem for the coefficients $C_k$, which is solved using the \texttt{SpectralB} Mathematica package~\cite{Fortuna:2020obg}. The method requires compactification of the infinite domain $z \in (-\infty, \infty)$ to $u \in [-1, 1]$ via a coordinate transformation, and careful handling of the outgoing wave boundary conditions. The calculated QNFs are listed in Tables~\ref{tab2}--\ref{tab4}, showing excellent agreement with other methods.

\begin{table}[htbp]
	\centering
	\begin{tabular}{|c|c|c|c|c|}
		\hline
		$s$  &
		$n$  &
		$\text{Bernstein spectral method}$  &
		$\;\text{Direct integration method}\;$ 	&
		$\;\;\;\text{Time evolution}\;\;\;$\\
		\hline
		~  &~   &$\text{Re}(m/k)$  ~~  $\text{Im}(m/k)$  &$\text{Re}(m/k)$ ~ $\text{Im}(m/k)$  &$\text{Re}(m/k)$ ~ $\text{Im}(m/k)$     \\
		1    &1   &1.99929~~ -0.50734            &~~1.99929 ~~ -0.50734               &~~1.99683~~~~   -0.50018\\
		&2   &1.74700~~ -1.57677                 &~~1.74655 ~~ -1.57582               &~~ non~~~~~~~~non \\
		3    &1   &1.83620 ~~ -0.12651           &~~1.83620 ~~ -0.12651               &~~1.83620~~~~   -0.12654\\
		&2   &2.75096~~ -0.61712                 &~~2.75096 ~~ -0.61712               &~~2.73050~~~~   -0.60744 \\
		5    &1   &1.80318~~ -0.09925            &~~1.80318  ~~ -0.09925              &~~1.80317~~~~   -0.09928\\
		&2   &2.79414~~ -0.43210                 &~~2.79414 ~~ -0.43210               &~~2.78422~~~~   -0.43857 \\ 
		7    &1   &1.79033~~ -0.09324            &~~1.79033 ~~ -0.09324               &~~1.79032~~~~   -0.09325\\
		&2   &2.80145~~ -0.37978                 &~~2.80145 ~~ -0.37978               &~~2.80018~~~~   -0.37935\\
		\hline
	\end{tabular}
	\caption{Low-overtone QNFs for $\delta=1$ and various $s$ values.\label{tab2}}
\end{table}

\begin{table*}[htbp]
	\centering
	\begin{tabular}{|c|c|c|c|}
		\hline
		$\;\;s\;\;$  &
		$\;\;n\;\;$  &
		$\;\;\text{Bernstein spectral method}\;\;$  &
		$\;\;\;\;\;\;\;\;\text{Asymptotic iteration method}\;\;\;\;\;\;\;$ 	\\
		\hline
		~  &~   &~~~~$\text{Re}(m/k)$  ~~  $\text{Im}(m/k)~~$  &$~~~~~~\text{Re}(m/k)$ ~~ $\text{Im}(m/k)~~$   \\
		~    &1   &1.99929~~ -0.50734           &~~1.99929 ~~  -0.50734                \\
		~    &2   &1.74700~~ -1.57677           &~~1.74700 ~~ -1.57677           \\
		~  	 &3   &1.27645~~ -2.86270           &~~1.27645 ~~ -2.86270           \\
		~    &4   &0.83946~~ -4.47669           &~~0.83946 ~~ -4.47670         \\
		1  	 &5   &0.60626~~ -6.19784           &~~0.60630 ~~ -6.19762              \\
		~    &6   &0.48644~~ -7.89599          &~~0.48578 ~~ -7.89518               \\
		~	 &7   &0.41627~~ -9.56403         &~~~0.41630~~ -9.56097              \\		
		~    &8   &~0.36995~~ -11.20640         &~~~0.36745~~ -11.21836             \\	
		~	 &9   &~0.33823~~ -12.84380         &~~~0.33437~~ -12.82631              \\		
		~    &10  &~0.30476~~ -14.46073         &~~~0.32597~~ -14.45580              \\		
		\hline
	\end{tabular}
	\caption{The first ten QNFs using the Bernstein spectral method and the asymptotic iteration method.\label{tab3}}
\end{table*}

\begin{table*}[htbp]
	\centering
	\begin{tabular}{|c|c|c|c|}
		\hline
		$\;\;s\;\;$  &
		$\;\;n\;\;$  &
		$\;\;\text{Bernstein spectral method}\;\;$  &
		$\;\;\;\;\;\;\;\;\text{Direct integration method}\;\;\;\;\;\;\;$ 	\\
		\hline
		~  &~   &~~~~$\text{Re}(m/k)$  ~~  $\text{Im}(m/k)~~$  &$~~~~~~\text{Re}(m/k)$ ~~ $\text{Im}(m/k)~~$   \\
		~    &1   &1.77953~~ -0.09065           &~~1.77953 ~~  -0.09065                \\
		~    &2   &2.80223~~ -0.35001           &~~2.80224 ~~ -0.35001           \\
		~    &3   &4.16862~~ -0.62184           &~~4.16864  ~~ -0.62172            \\
		~    &4   &5.63090~~ -0.87406           &~~5.63111 ~~ -0.87389         \\
		11   &5    &7.13462~~ -1.11123           &~~7.13407 ~~ -1.11181               \\
		~    &6     &8.65587~~ -1.34219          &~~8.65662~~ -1.34296              \\
		~    &7     &10.18781~~ -1.56976          &~~10.18986~~ -1.57089           \\		
		~    &8     &11.72844~~ -1.79461          &~~11.72942~~ -1.79721           \\	
		~    &9     &13.27754~~ -2.02014          &~~13.27291~~ -2.02266          \\
		~    &10    & 14.81845~~  -2.25382        &~~14.81900~~ -2.24761         \\
		\hline
		~    &1   &1.77234~~   -0.09050        &~~1.77234~~  -0.09048               \\
		~    &2   &2.79864~~ -0.33932          &~~2.79898~~  -0.33986           \\
		~  	 &3   &4.18069~~ -0.58177          &~~4.18059~~  -0.57604            \\
		~    &4   &5.65238~~ -0.77869          &~~5.65775~~  -0.77201          \\
		19 	 &5   &7.18011~~ -0.92683          &~~7.17339~~  -0.94005                \\
		~    &6   &8.69511~~ -1.08414          &~~8.70706~~  -1.09238               \\
		~	 &7   &10.22848~~ -1.22761         &~~10.25055~~ -1.23573              \\		
		~    &8   &11.78923~~ -1.35713         &~~11.80000~~ -1.37382              \\	
		~	 &9   &13.37321~~ -1.48809         &~~13.35333~~ -1.50879            \\
		~    &10  &14.91535~~ -1.67253         &~~14.90934~~ -1.64188                  \\
		\hline
	\end{tabular}
	\caption{The first ten QNFs using the Bernstein spectral method and the direct integration method.\label{tab4}}
\end{table*}

\subsubsection{Case 3: $s > 1, \delta>1$}
\begin{figure}
	\centering
	\includegraphics[width=0.4\textwidth]{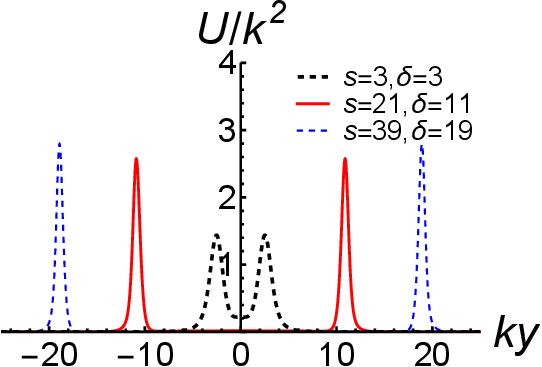}
	\caption{Plots of the effective potential $U(y)$ for the case 3.}\label{figeffectivePcase3}
\end{figure}

This regime presents the greatest complexity: the $y$-$z$ relationship is unknown, and the effective potential exhibits a pronounced double-barrier structure (Fig.~\ref{figeffectivePcase3}), which is a prerequisite for echo phenomena. The Bernstein and WKB methods are inapplicable here. We rely on the direct integration method to compute the QNFs, results of which are presented in Table~\ref{tab5}. This case is particularly significant as it is the most promising regime for observing gravitational echoes.

\begin{table}[htbp]
	\centering
	\begin{tabular}{|c|c|c|c|c|}
		\hline
		$s$  &
		$\delta$  &
		$n$  &
		$\;\text{Direct integration method}\;$ 	&
		$\;\;\;\text{Time evolution}\;\;\;$\\
		\hline
		~  &~   &~ &$\text{Re}(m/k)$ ~ $\text{Im}(m/k)$  &$\text{Re}(m/k)$ ~ $\text{Im}(m/k)$     \\
		3    &3   &1      &~~0.75397 ~~ -0.00738              &~~0.75397~~~~   -0.00738\\
	         &   &2      &~~1.17777 ~~ -0.07748                &~~1.17777~~~~  -0.07748\\
	         &   &3      &~~1.57294 ~~ -0.26796                &~~ non~~~~~~~~non \\
	         &   &4      &~~1.99265 ~~ -0.53685                &~~ non~~~~~~~~non \\
		21   &11   &1      &~~0.18877 ~~-1.66$\times 10^{-5}$               &~~ non~~~~~~~~non \\
		     &    &2            &~~0.31006 ~~ -1.05$\times 10^{-4}$              &~~ non~~~~~~~~non \\
		     &    &3            &~~0.44645 ~~ -3.39$\times 10^{-4}$              &~~ non~~~~~~~~non \\
		     &    &4           &~~0.58562 ~~ -8.15$\times 10^{-4}$                &~~ non~~~~~~~~non \\
		39    &19   &1      &~~0.10892  ~~ -1.96$\times 10^{-6}$               &~~ non~~~~~~~~non \\
		&    &2           &~~0.17875 ~~ -1.22$\times 10^{-5}$              &~~ non~~~~~~~~non \\ 
		     &    &3      &~~0.25756 ~~ -3.85$\times 10^{-5}$              &~~ non~~~~~~~~non \\
		&    &4           &~~0.33827 ~~ -8.83$\times 10^{-5}$               &~~ non~~~~~~~~non \\
		\hline
	\end{tabular}
	\caption{Low-overtone QNFs for Case 3 ($s > 1, \delta > 1$) obtained via direct integration. ``non'' indicates modes not clearly extractable from time evolution, likely due to echo interference.\label{tab5}}
\end{table}

\subsection{Time domain: waveforms of QNMs and echoes}
To complement the frequency-domain analysis and gain insight into the temporal dynamics and excitation amplitudes of the QNMs, we perform numerical time evolution of initial wave packets. We solve the evolution equation~\eqref{evolutionequation} in light-cone coordinates $(u = t-z, v = t+z)$, where it takes the form:
\begin{eqnarray}
	\left(4\frac{\partial^{2}}{\partial u\partial v}+U+a^{2}\right)\Phi(u,v)=0. \label{uvevolutionequation}
\end{eqnarray}
We consider two types of initial data:
\begin{enumerate}
	\item A Gaussian wave packet, centered at $v_c$ with width $\sigma$:
	\begin{equation}
		\Phi(0, v) = \exp\left(-\frac{(v - v_c)^2}{2\sigma^2}\right), \quad \Phi(u, 0) = \exp\left(-\frac{v_c^2}{2\sigma^2}\right).
		\label{gausspulseinitialwavepackage}
	\end{equation}
	This excites all modes, but the first mode dominates the response.
	\item A static, odd-parity wave packet:
	\begin{equation}
		\Phi(0, v) = \sin\left(\frac{k v}{2}\right) \exp\left(-\frac{k^2 v^2}{4}\right), \quad \Phi(u, 0) = \sin\left(\frac{k u}{2}\right) \exp\left(-\frac{k^2 u^2}{4}\right).
		\label{oddinitialwavepackage}
	\end{equation}
	This selectively excites odd-parity modes, enabling the study of higher overtones.
\end{enumerate}

\subsubsection{Case 1: $s=1$, $\delta \geq 1$}
\begin{figure}
	\centering
	\subfigure[~Gauss packet]{\label{figqnmsplotcase1even}
		\includegraphics[width=0.4\textwidth]{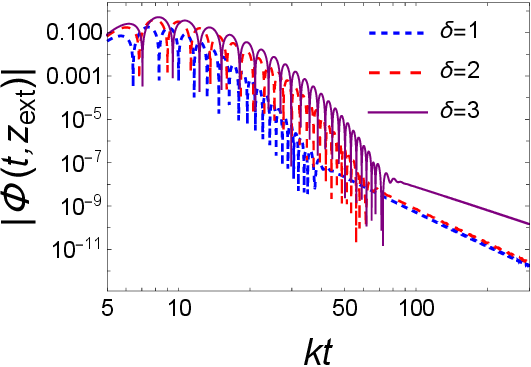}}
	\subfigure[~static odd parity packet]{\label{figqnmsplotcase1odd}
		\includegraphics[width=0.4\textwidth]{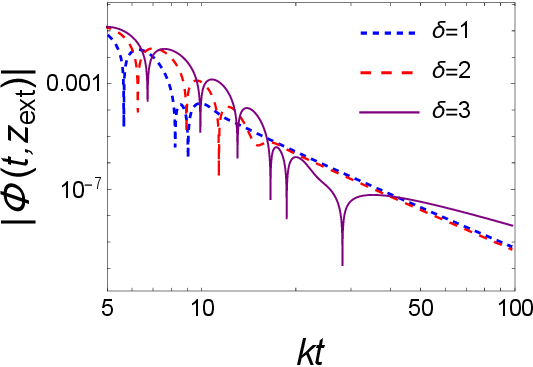}}
	\caption{Time evolution of wave packets in Case 1 ($s=1, \delta\geq 1$), plotted on a logarithmic scale.}\label{figcase1}
\end{figure}

The evolution of a Gaussian packet (Fig.~\ref{figcase1}, left) displays the canonical QNM ringing: an initial burst, followed by an exponential decay of oscillations (the QNM phase), and a late-time power-law tail. The extracted frequencies (Tab.~\ref{tab1}) agree well with frequency-domain results. Increasing $\delta$ decreases the oscillation frequency and increases the lifetime. The evolution of the odd packet (Fig.~\ref{figcase1}, right) confirms the existence and properties of the first odd overtone.

\subsubsection{Case 2: $\delta=1$, $s \geq 1$}
\begin{figure}
	\centering
	\subfigure[~Gauss packet]{\label{figqnmsplotcase2even}
		\includegraphics[width=0.4\textwidth]{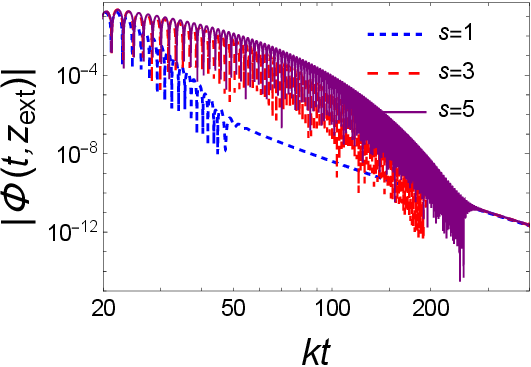}}
	\subfigure[~static odd parity packet]{\label{figqnmsplotcase2odd}
		\includegraphics[width=0.4\textwidth]{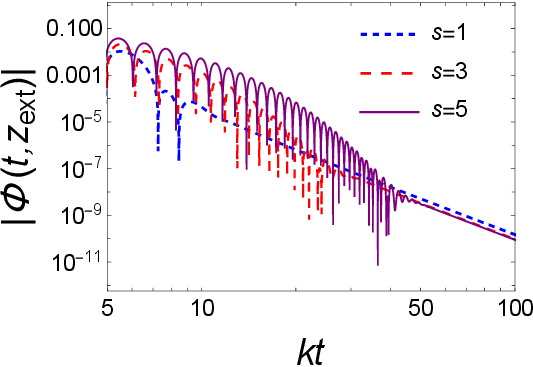}}
	\caption{Time evolution of wave packets in Case 2 ($s\geq1, \delta=1$), plotted on a logarithmic scale.}\label{figcase2}
\end{figure}

The split potential significantly prolongs the QNM lifetime for both even and odd parities (Fig.~\ref{figcase2}). The extracted frequencies (Tab.~\ref{tab2}) again validate the frequency-domain results. The late-time tail remains a power law, consistent with the asymptotic structure of the potential. No echoes are observed for these parameters.

\subsubsection{Case 3: $s > 1$, $\delta > 1$}
\begin{figure}
	\centering
	\subfigure[~$s=\delta=3$]{\label{figqnmsplotcase3s3d3z0}
		\includegraphics[width=0.4\textwidth]{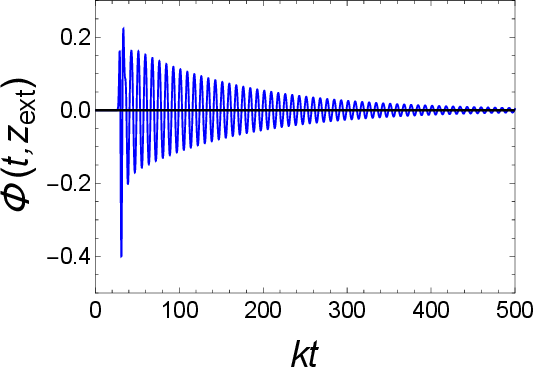}}
	\subfigure[~$s=\delta=3$]{\label{figqnmsplotlogcase3s3d3z0}
		\includegraphics[width=0.4\textwidth]{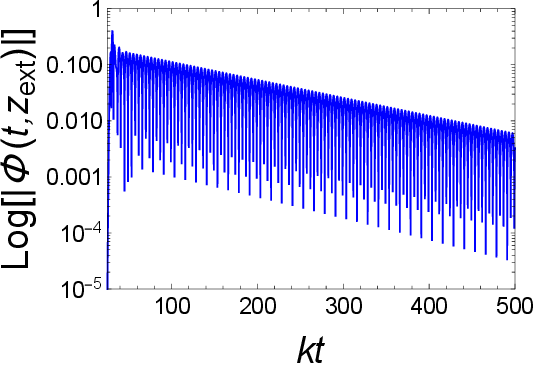}}
			\subfigure[~$s=21, \delta=11$]{\label{figqnmsplotcase3s21d11z10}
			\includegraphics[width=0.4\textwidth]{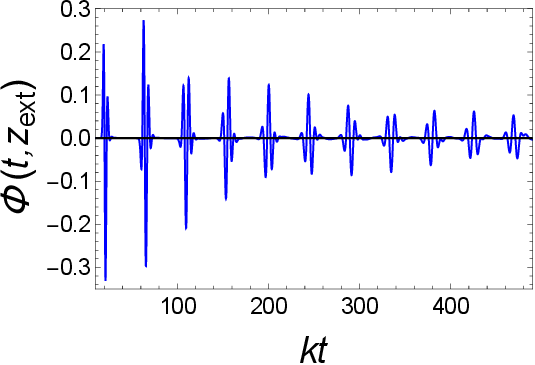}}
		\subfigure[~$s=21, \delta=11$]{\label{figqnmsplotlogcase3s21d11z10}
			\includegraphics[width=0.4\textwidth]{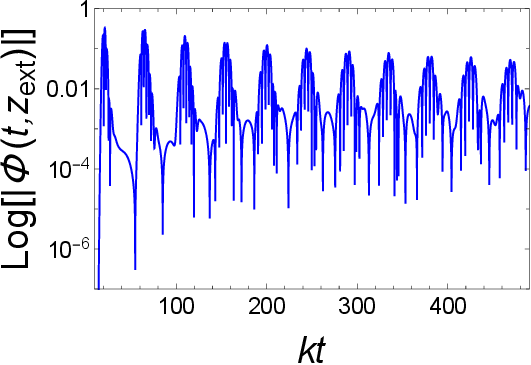}}
	\caption{Left panel: Evolution waveforms of the Gauss wave packet in Case 3. Right panel: Same as the left panel but plotted on a semi-logarithmic scale.}\label{figcase31}
\end{figure}

\begin{figure}
	\centering
	\subfigure[~$kz_{\text{ext}}=10$]{\label{figqnmsplotcase3s39d19z10}
		\includegraphics[width=0.4\textwidth]{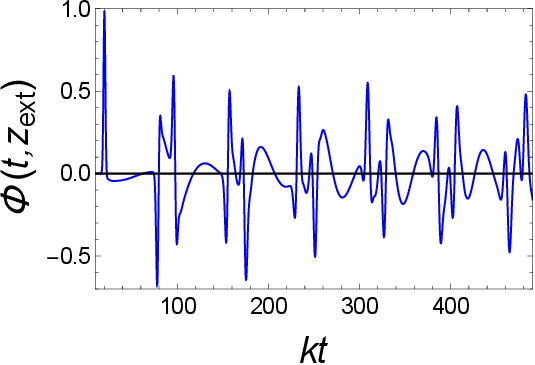}}
	\subfigure[~$kz_{\text{ext}}=10$]{\label{figqnmsplotlogcase3s39d19z10}
		\includegraphics[width=0.4\textwidth]{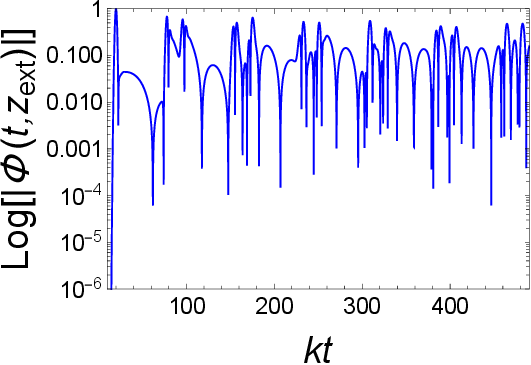}}
	\subfigure[~$kz_{\text{ext}}=20$]{\label{figqnmsplotcase3s39d19z20}
		\includegraphics[width=0.4\textwidth]{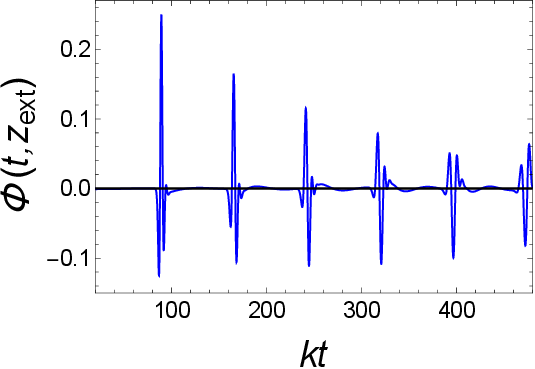}}
	\subfigure[~$kz_{\text{ext}}=20$]{\label{figqnmsplotlogcase3s39d19z20}
		\includegraphics[width=0.4\textwidth]{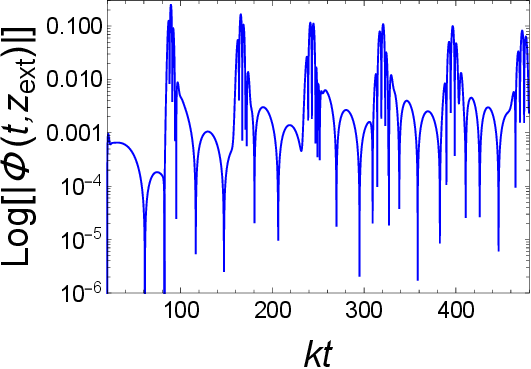}}
	\caption{Left panel: Evolution waveforms of the Gauss wave packet at different observation points, the parameters are set as $s=39, \delta=19$. Right panel: Same as the left panel but plotted on a semi-logarithmic scale.}\label{figcase32}
\end{figure}

This is the key regime for echo discovery. For smaller $\delta$ and $s$ (e.g., $s=\delta=3$, Fig.~\ref{figcase31}a,b), only long-lived QNMs are observed. For larger values (e.g., $s=21, \delta=11$, Fig.~\ref{figcase31}c,d), clear, periodic echo signals emerge after the initial QNM ringdown. The echo waveform is highly dependent on the observer's position relative to the sub-branes (Fig.~\ref{figcase32}). An observer on a sub-brane sees a clean echo period set by the inter-brane distance. An observer between the sub-branes sees a more complex signal featuring a second periodicity related to their intra-brane position. This intriguing property could potentially be used to infer the relative position within the brane structure.

\section{Conclusion}\label{sec:conclusions}

In this work, we have conducted a comprehensive investigation of the graviscalar quasinormal modes (QNMs) and the phenomenon of gravitational echoes within a thick brane-world scenario described by a canonical scalar field. Our analysis, performed in both the frequency and time domains, provides a detailed characterization of the spectrum of fluctuations and their dynamics, particularly in the context of branes that split into sub-structures.

The frequency-domain study, employing the WKB approximation, the direct integration method, and the Bernstein spectral method, successfully determined the complex quasinormal frequencies across three distinct parametric regimes. For the single, non-splitting brane ($s=1$), the QNM spectrum was found to consist of a series of modes whose oscillation frequencies decrease and lifetimes increase with the parameter $\delta$, which controls the brane's thickness. The excellent agreement between the WKB and direct integration methods validated these results for the fundamental and first overtone modes. In the case of a split brane with $\delta=1$ and $s>1$, the Bernstein spectral method proved essential for accurately calculating higher overtones, revealing a spectrum that is significantly influenced by the double-barrier structure of the effective potential. Finally, for the most general case ($s > 1, \delta > 1$), where the potential exhibits a pronounced double barrier and the coordinate transformation is intractable, the direct integration method was indispensable. The calculated QNFs in this regime show markedly longer lifetimes, a prerequisite for the emergence of late-time echoes.

The time-domain evolution of Gaussian and specialized odd-parity wave packets not only confirmed the frequency-domain results but also provided crucial insights into the temporal dynamics and excitation of these modes. The characteristic ringdown phase, followed by a power-law tail, was clearly observed. Most significantly, for sufficiently large $s$ and $\delta$ parameters, our simulations unveiled clear and persistent gravitational echo signals. These echoes result from the successive reflections of the waveform between the two potential barriers created by the split brane. A particularly novel finding is the dependence of the echo waveform on the observer's position within the extra dimension. An observer located on a sub-brane observes a clean echo signal with a period dictated solely by the distance between the sub-branes. In contrast, an observer situated between the sub-branes detects a more complex signal featuring a superposition of periodicities, hinting at the potential to infer the relative position of an observer within the brane structure.

In conclusion, our study establishes that thick brane models with internal structure, parameterized by $s$ and $\delta$, support a rich spectrum of graviscalar quasinormal modes and can produce detectable gravitational echoes. The synergy between multiple analytical methods in the frequency domain and numerical evolution in the time domain has been crucial for a robust characterization of these phenomena. Based on the similarity between the effective potential of our braneworld scenario and those found in black hole/wormhole geometries, our findings provide valuable insights and methodological approaches for investigating echo phenomena in these compact objects. The observation of position-dependent echo signatures presents a unique and intriguing observational fingerprint for braneworld scenarios with sub-structure. Future work could involve a more extensive parameter survey and the exploration of echoes in other types of brane models, potentially forging a link between fundamental theory and observable gravitational wave signals.

\acknowledgments
This work was supported by the National Natural Science Foundation of China (Grants No. 12035005, No. 12405055, and No. 12347111), the China Postdoctoral Science Foundation (Grant No. 2023M741148), the Postdoctoral Fellowship Program of CPSF (Grant No. GZC20240458), and the National Key Research and Development Program of China (Grant No. 2020YFC2201400).


\end{document}